\documentclass[aps, twocolumn, 10pt, showpacs, longbibliography, superscriptaddress]{revtex4-1}
\usepackage{fullpage}
\usepackage{graphicx}
\usepackage{dcolumn}
\usepackage{bm}
\usepackage{color}
\usepackage{amsmath, amsthm, amssymb}
\usepackage[modulo]{lineno}
\usepackage[usenames,dvipsnames]{xcolor}
\usepackage{ulem}
\usepackage{epstopdf}
\newcommand{\ba}{\begin{eqnarray}}
\newcommand{\ea}{\end{eqnarray}}

\usepackage{float}
\floatstyle{boxed}

\graphicspath{ {Figures/} }

\begin{document}
\title{Graphene-based Josephson junction single photon detector}
\author{Evan D. Walsh}
\affiliation{Department of Electrical Engineering and Computer Science, Massachusetts Institute of Technology, Cambridge, MA 02139}
\affiliation{School of Engineering and Applied Sciences, Harvard University, Cambridge, MA 02138}
\author{Dmitri K. Efetov}
\affiliation{Department of Electrical Engineering and Computer Science, Massachusetts Institute of Technology, Cambridge, MA 02139}
\affiliation{ICFO-Institut de Ci\`{e}ncies Fot\`{o}niques, The Barcelona Institute of Science and Technology, 08860 Castelldefels (Barcelona), Spain}
\author{Gil-Ho Lee}
\affiliation{Department of Physics, Harvard University, Cambridge, MA 02138}
\author{Mikkel Heuck}
\affiliation{Department of Electrical Engineering and Computer Science, Massachusetts Institute of Technology, Cambridge, MA 02139}
\author{Jesse Crossno}
\affiliation{School of Engineering and Applied Sciences, Harvard University, Cambridge, MA 02138}
\author{Thomas A. Ohki}
\affiliation{Raytheon BBN Technologies, Quantum Information Processing Group, Cambridge, Massachusetts 02138, USA}
\author{Philip Kim}
\affiliation{Department of Physics, Harvard University, Cambridge, MA 02138}
\author{Dirk Englund}
\email{englund@mit.edu}
\affiliation{Department of Electrical Engineering and Computer Science, Massachusetts Institute of Technology, Cambridge, MA 02139}
\author{Kin Chung Fong}
\email{kc.fong@raytheon.com}
\affiliation{Raytheon BBN Technologies, Quantum Information Processing Group, Cambridge, Massachusetts 02138, USA}
\date{\today}

\begin{abstract}
We propose to use graphene-based Josephson junctions (gJjs) to detect single photons in a wide electromagnetic spectrum from visible to radio frequencies. Our approach takes advantage of the exceptionally low electronic heat capacity of monolayer graphene and its constricted thermal conductance to its phonon degrees of freedom. Such a system could provide high sensitivity photon detection required for research areas including quantum information processing and radio-astronomy. As an example, we present our device concepts for gJj single photon detectors in both the microwave and infrared regimes. The dark count rate and intrinsic quantum efficiency are computed based on parameters from a measured gJj, demonstrating feasibility within existing technologies.
\end{abstract}
\maketitle

\section{Introduction}
\label{Intro}
Detecting single light quanta enables technologies across a wide electromagnetic (EM) spectrum. In the infrared regime, single photon detectors (SPD) are essential components for deep space optical communication \cite{Shaw:15} and quantum key distribution via fiber networks \cite{Takesue:2007cya}. In frequencies on the order of terahertz, single photon detectors will allow the study of galaxy formation through the cosmic infrared background with an estimated photon flux $<$100 Hz \cite{Wei:2008jw}. Microwave SPDs and photon number resolving counters are required in a number of proposed quantum technologies, including remote entanglement of superconducting qubits \cite{Narla:2016bh}, high-fidelity quantum measurements \cite{Romero:2009ek}, and microwave quantum illumination \cite{Barzanjeh:2015id}. However, detecting low frequency photons is challenging because of their vanishingly small energy. A prominently used detection scheme is to exploit the heating effect from single photons. For instance, transition edge sensors and superconducting nanowire single photon detectors can register infrared photons as they break Cooper pairs in the superconductors \cite{Hadfield:2009es}. High sensitivity calorimeters can detect single photons by reading out the temperature rise induced by the absorbed photon but require better heat absorbers to reach single photon sensitivity at lower frequencies \cite{Gasparinetti:2015gr}.

Graphene is a promising material for single photon calorimetry \cite{Fong:2012ut, McKitterick:2013ue, Yan:2012eh}. With its pseudo-relativistic band structure, graphene can efficiently absorb photons from a wide EM spectrum, making it attractive for expanding the availability of SPDs to applications in a broader frequency range. Compared to metals, the electron-phonon coupling and electronic specific heat capacity of monolayer graphene are extremely small \cite{Fong:2012ut, Vora:2012cs} due to the shrinking density of states near its charge neutrality point (CNP). Therefore, a single photon absorbed by graphene can heat up the electrons significantly. This approach relies on thermal physics in this extraordinary material in contrast to atomic-like systems such as quantum dots and superconducting circuits \cite{Chen:2011fj, Inomata:2016jc} which require more complicated operation protocols such as microwave pumping before detecting photons.

Sensing the heat pulse generated from a single photon can be challenging experimentally. Although noise thermometry may have the bandwidth and sensitivity to read out the temperature rise, this rise of electron temperature also degrades its temperature resolution which can translate to poor dark count characteristics \cite{McKitterick:2013ue}. Here we propose using the graphene-based superconducting-normal-superconducting (SNS) Josephson junction (Jj) as a threshold sensor to detect single photons across an extremely wide spectrum. Since the first observation of the superconducting proximity effect in graphene \cite{Heersche:2007}, many advances have been made in the fabrication and performance of graphene-based Jjs (gJjs) \cite{Lee:2011et, Jeong:2011dp, Mizuno:2013dz, Calado:2015fp, BenShalom:2015hy, Borzenets:2016gw}. To emphasize feasibility with existing materials and fabrication technologies, we used measured parameters from a gJj to calculate the performance of our proposed SPD. Our modeling suggests that a low dark count probability with an intrinsic quantum efficiency approaching unity is achievable. 

\section{Device concept and Input coupling}

\begin{figure}[t]
\includegraphics[width=0.5\columnwidth]{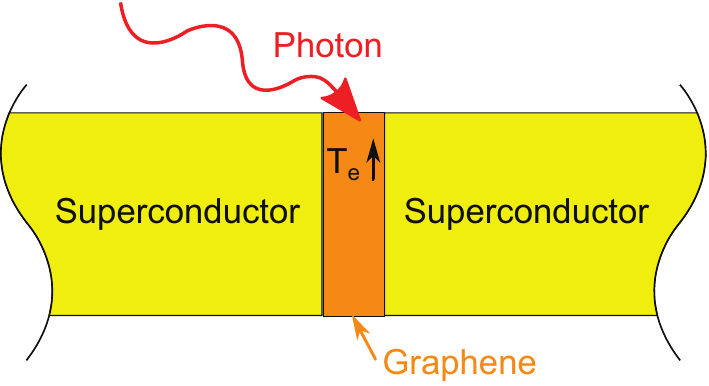}
\caption{Device concept to detect single photon using a graphene-based Josephson junction.}
\label{figDeviceConcept}
\end{figure}

Our proposed device is a hybrid of the calorimeter and the SNS Jj. SNS Jjs have been recognized for their use as superconducting transistors \cite{Wilhelm:1998hk} and as sensitive bolometers\cite{Giazotto:2008, Voutilainen:2010, Govenius:2016}. The concept is to achieve control of the supercurrent by perturbing the Fermi distribution of the normal constituent in the junction through joule heating \cite{Morpurgo:1998bp}. Compared to metals and semiconductors, graphene is a more favorable weak link material with its high electronic mobility, sensitive thermal response, and field-tunable chemical potential. When an absorbed photon raises the electron temperature in the graphene sheet, the calorimetric effect can trigger the Jj to switch from the zero voltage to resistive state (see Fig. 1). We can describe this heating with a quasi-equilibrium temperature, $T_e$, of the graphene electrons as they thermalize quickly through electron-electron interactions \cite{Tielrooij:2013cd, Brida:2013hi}. A gJj SPD can be achieved by efficient photon absorption (discussed in the present section), appreciable temperature elevation (Section \ref{sectionGraphene}), and sensitive transition of the gJj (Section \ref{sectionGJJ}).

\begin{figure*}[t]
\includegraphics[width=2\columnwidth]{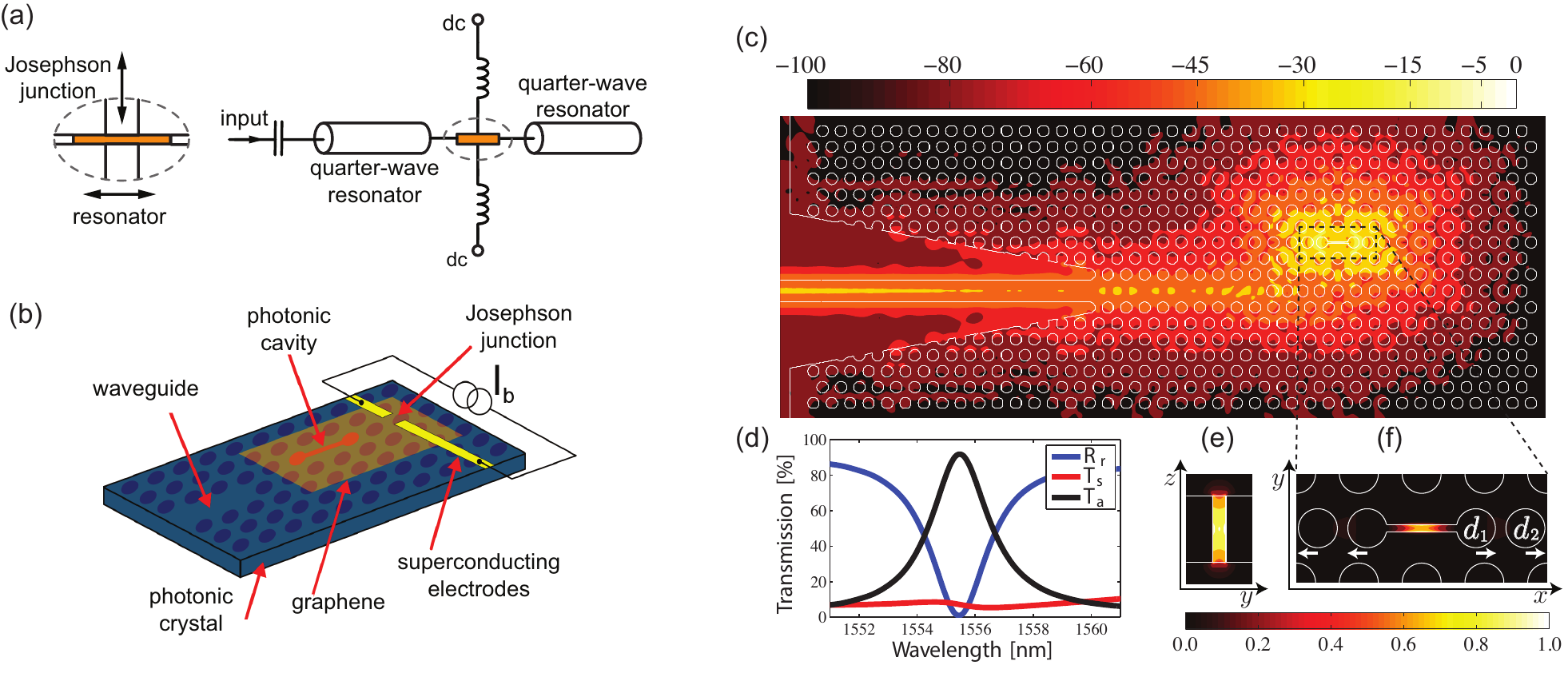}
\caption{Device schematic for (a) microwave and (b) infrared single photon detection. (a) The graphene flake is located at the current antinode of a halfwave microwave resonator for maximizing input efficiency. Two stages of inductors and capacitors form a high impedance network at microwave frequency for the dc measurement of the gJj. (b) A graphene sheet lies on top of a PhC cavity to increase its absorption through critical coupling. Light is coupled into the cavity through an in-chip waveguide. (c) Simulation results for critical coupling. Top view of the PhC structure overlaid with the mode profile, $|\bm{E}|^2$, using a dB scale normalized to a maximum of 1. (d) Spectra showing the wavelength dependence of the reflection, $R_r$, absorption, $T_a$, and scattered power, $T_s$. (e) Cross-sectional and (f) planar view of close-ups of the cavity mode, $|\bm{E}|^2$, using a linear scale. The parameters of the structure are: Membrane thickness, $h$ = 250 nm, lattice period, $a$ = 0.27$\lambda$, hole radius, $r = 0.31a$, cavity slot width, $w_s$ = 0.032$\lambda$, cavity hole shifts, $d_1 = 0.365a$, and $d_2 = 0.153a$, waveguide width, $w_w = 2(W\sqrt{3}a/2-r)$ where $W$ = 1.04. The holes at the termination of the PhC waveguide are shifted along the $x$-axis by $s_1 = 0.44a$, $s_2 = 0.27a$, and $s_3=0.1a$.}
\label{figCoupling}
\end{figure*}

A practical challenge concerns the efficient absorption of EM radiation by the graphene sheet. Using Boltzmann transport, the high frequency conductivity of monolayer graphene can be described in a Drude form \cite{Horng:2011cr}. At low frequencies, graphene is essentially a lump resistor with impedance depending on its dimensions and gate-tunable charge carrier density. Quarter or half-wave resonators, stub or LC matching networks \cite{Fong:2012ut}, taper transformers, and log periodic antennas \cite{Vicarelli:2012de} can be employed to achieve an input coupling approaching unity. Broadband 50 $\Omega$ matching is achievable with highly doped monolayer graphene. At optical frequencies, normal incidence light absorption is given by $\pi\alpha\simeq2\%$ where $\alpha$ is the fine structure constant, due to the universal a.c. conductivity \cite{Gusynin:2006fg}. However, EM waves can be absorbed efficiently by evanescent wave coupling, with light grazing across the graphene sheet, using waveguides and photonic crystal (PhC) structures \cite{Gan:2012eu}. Fig.~\ref{figCoupling}A and B illustrate the proposed gJj SPDs using impedance-matched resonators at microwave and infrared frequencies, respectively.

To couple microwave photons and apply a dc current bias simultaneously, the gJj is embedded in a four-terminal geometry as shown in Fig.~\ref{figCoupling}A. Supercurrent flows from the narrowly gapped (vertical direction in the figure) superconducting contacts through the monolayer graphene (orange) by the proximity effect. The inductive chokes following these Jj contacts isolate the microwave coupling and permit fast Jj switching. For microwave operation, one quarter-wave microwave resonator is in contact on each side of the graphene sheet along the direction of wider separation between the superconducting terminals (horizontal direction in the figure). Together, they form a half-wave resonator with the dissipative graphene sheet at the microwave current antinode. High microwave absorption can be achieved by impedance matching the half-wave resonator while the temporal mode of the single photon determines the optimal quality factor for single photon detection \cite{Pechal:2014db}.

For infrared photodetection, a dielectric photonic crystal cavity can provide the impedance-matching element to reach near-unity light absorption by the graphene sheet, as illustrated in Fig.~\ref{figCoupling}B. Infrared radiation passes from a ridge waveguide into a PhC nanocavity, via a short section of PhC waveguide. The evanescent cavity field couples to the graphene monolayer, positioned over the cavity, while the Jj is located at the other end. Graphene can be critically coupled to the cavity \cite{Gan:2012eu} so that all incident light is absorbed. Our PhC cavity design uses a thin air slot to concentrate the EM field into the graphene sheet; this air-slot (and the graphene absorber) can be deeply sub-wavelength. Fig.~\ref{figCoupling}C shows the EM field concentration into the PhC cavity air-slot. Using a finite-difference time-domain simulation tool (Lumerical), we calculate the reflected, absorbed, and scattered power for a broadband optical input pulse from the ridge waveguide (plotted in Fig. 2d). For a graphene-cavity quality factor of 800, the calculated power spectrum indicates a peak graphene absorption of 93\% for 1550 nm wavelength photons. Since the optical losses in silicon are comparatively negligible, the remaining losses are due to optical scattering, which can likely be eliminated by further numerical optimization. 

\section{Graphene thermal response}
\label{sectionGraphene}

Upon absorbing a single photon, the thermal response of the graphene electrons can be characterized by the thermal time constant $\tau_{th}$, heat capacity $C_e$, and thermal conductance $G_{th}$ to the reservoir. Due to the fast electron-electron interaction time, the photon energy can quickly thermalize among the graphene electrons and establish a quasi-equilibrium in typically tens of femtoseconds \cite{Tielrooij:2013cd, Brida:2013hi}. Therefore, both the heat injection from the photon and the initial temperature rise can be considered instantaneous when compared to the thermal time constant of the graphene electrons. 

This initial temperature rise is determined using the electronic heat capacity of the monolayer graphene. In the degeneracy regime, \ba C_e = A\gamma T\label{EqnCe}\ea \cite{Viljas:2010jj}, where $A$ is the area of the graphene sheet and $\gamma=(4\pi^{5/2}k_B^2n^{1/2})/(3hv_F)$ is the Sommerfeld coefficient with $k_B$ and $h$ being Boltzmann's and Planck's constants, respectively. This is in contrast to non-degenerate Dirac fermions where $C_e\propto T^2$ \cite{Vafek:2007ce}. In graphene, the electron Fermi energy $E_F$ has a Dirac-like dispersion relation, i.e. $E_F= \hbar v_F k_F$, where $\hbar = h/2\pi$, $v_F=10^6$ ms$^{-1}$ is the graphene Fermi velocity, and $k_F = \sqrt{\pi n}$ is the Fermi momentum with $n$ being the charge carrier density. Thus for the typical charge carrier density ranging from $10^{10}$ to $10^{12}$ cm$^{-2}$, $E_F$ is higher than 10 m$e$V so that the Fermi temperature is about 135 K, justifying the use of Eqn. (\ref{EqnCe}). $C_e$ can be tuned using a gate voltage reaching a minimum at the charge neutrality point, where it is limited by the residual puddle density \cite{Wang:2013ch}. We plot $C_e$ in Fig. 3a at a carrier density $n_0 = 1.7\times 10^{12}$ cm$^{-2}$ that we will use in the modeling. Compared to that of a metallic nanowire used in photon detection at the same temperature \cite{Wei:2008jw}, the electronic heat capacity of a graphene sheet of 1 $\mu$m$^2$ area would be more than three orders of magnitude smaller. This dramatic improvement is due to the shrinking density of states, $D(E)=2|E|A/\pi(\hbar v_F)^2$ with $E$ being the energy measured from the CNP, in monolayer graphene.

\begin{figure*}[t]
\includegraphics[width=2\columnwidth]{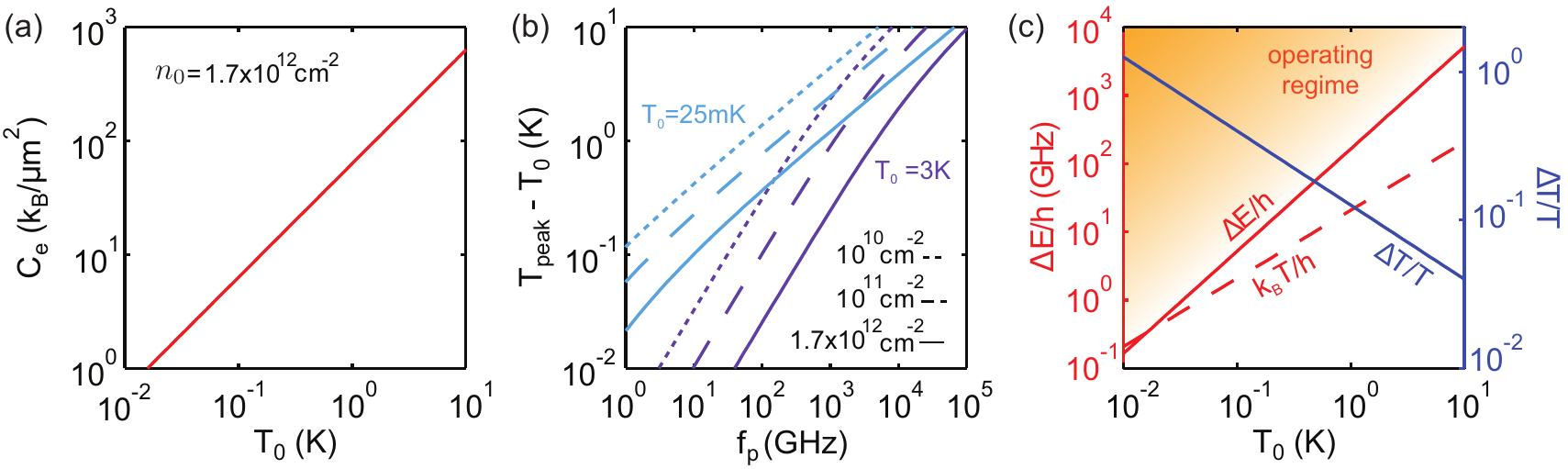}
\caption{(a) The specific heat of a 1$\mu$m$^2$ graphene sheet as a function of base temperature. (b) The initial temperature rise vs. frequency with electron density $n_0 = 1.7\times10^{12}$ cm$^{-2}$ (solid, used for modeling), 10$^{11}$ cm$^{-2}$ (dashed), or 10$^{10}$ cm$^{-2}$ (dotted) at a base temperature of 25 mK (blue) or 3 K (purple). For an infrared detector $T_0$ = 3 K will suffice but for microwave detection a lower $T_0$ is required to acquire a noticeable temperature change. (c) Energy resolution and temperature fluctuation of a 1 $\mu$m$^2$ graphene sheet at $n_0$, representing the intrinsic noise of the calorimeter.}
\label{figThermalResponseA}
\end{figure*}

We can estimate the initial temperature $T_{peak}$ of the hot electrons by equating the integrated internal energy $\int_{T_0}^{T_{peak}}C_e(T)dT$ to the photon energy such that \ba T_{peak} = \sqrt{(2 h f_p)/(\gamma A)+T_0^2}\ea \cite{McKitterick:2013ue}, where $T_0$ is the base temperature and $f_p$ is the photon frequency. Fig. 3b plots the temperature rise for various photon frequencies and charge carrier densities at 0.025 and 3 K. The temperature rise is higher for a lower charge carrier density or with a higher energy photon. Here we assume a full conversion of the photon energy to internal energy in the graphene electrons. This assumption is justified by pump-probe experiments from which it was inferred that up to 80\% of absorbed photon energy is cascaded down to heat electrons \cite{Tielrooij:2013cd}. This efficient energy conversion is due to the domination of the electron-electron scattering process over the coupling to the optical phonons. For lower energy photons at microwave frequencies, the heat leakage to optical phonons is negligible as the energy scale is further below the optical phonon energy.

The heat capacity also determines the root mean square fluctuations in energy of the graphene sheet, $\Delta E$, shown in Fig. 3c for $n=n_0$. This intrinsic noise of the calorimeter is given by \cite{Chui:1992wf} \ba\Delta E = \sqrt{C_ek_BT^2}\ea and describes the thermodynamic fluctuations of the electrons in graphene as a canonical ensemble in thermal equilibrium with a reservoir. $\Delta E$ sets the SPD energy resolution for a measurement time much longer than $\tau_{th}$. The fluctuation power spectral density at spectral frequency $f$ rolls off as $\sqrt{4\tau_{th}/(1+4\pi^2\tau_{th}^2f^2)}$ since $\tau_{th}$ determines the time scale of the energy exchange between the ensemble and reservoir. In principle, widening the measurement bandwidth $B$ can allow detection of the sharp temperature increase due to a single photon thus circumventing the limitations of calorimetry imposed by these intrinsic fluctuations \cite{McCammon}. However, it can also expose the Jj to high frequency noise and increase the thermal conductance of the radiation channel. In this report, we shall focus on the small bandwidth regime in which the SPD requires photon energy larger than $\Delta E$. The color gradient orange region in Fig. 3c highlights the requirement of operating temperature for a given photon frequency to avoid both the energy fluctuation and thermal noise. Related to the energy fluctuations are temperature fluctuations with root mean square $\Delta T$ given by: \ba \Delta T= \Delta E/C_e=\sqrt{k_BT^2/C_e}.\ea Curiously, when $C_e$ decreases to $k_B$, $\Delta T/T$ approaches unity (Fig. 3c). Possible modification of Boltzmann-Gibb statistics to describe the fluctuations when the degrees of freedom in the system are close to one is beyond the scope of this report \cite{Wilk:2000jx}. However, temperature fluctuations can affect Jj transitions and will be included in the performance calculation in Section \ref{sectionPerformance}.

\begin{figure*}[t]
\includegraphics[width=1.4\columnwidth]{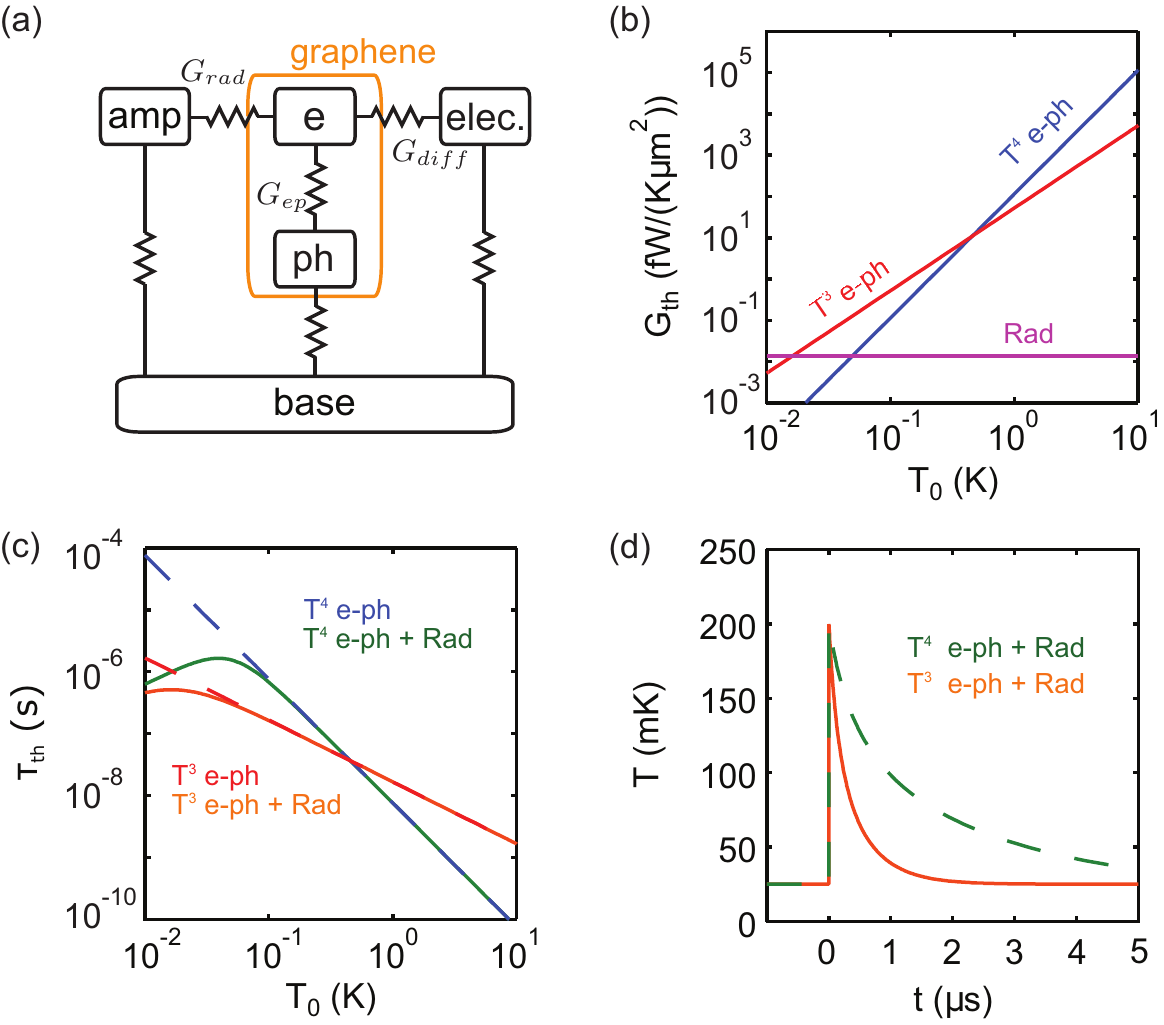}
\caption{(a) Thermal diagram depicting three heat transfer pathways from graphene electrons (e.) to the thermal reservoir (base) via graphene electron-phonon coupling (ph.), to the electrical contacts (elec.) via diffusion, or to the electrical environment, such as an amplifier (amp.), via photon emission (rad.). The thermal conductances in the linear response regime are denoted as $G_{ep}$, $G_{diff}$, and $G_{rad}$, correspondingly. (b) The thermal conductance vs. base temperature for clean graphene with a $T^4$ e-ph coupling law (blue), for disordered graphene with a $T^3$ e-ph coupling law (red), and for the radiation channel (purple). (c) The thermal time constant $\tau_{th}=C_e/G_{th}$ in the linear response regime for clean (green (blue) including (excluding) radiation) and disordered (orange (red) including (excluding) radiation) graphene. (d) The transient thermal response of the graphene sheet upon absorbing a 26 GHz microwave photon for clean (green) and disordered (orange) graphene including radiation.}
\label{figThermalResponseB}
\end{figure*}

Fig. 4a depicts the thermal pathways of a graphene sheet \cite{Fong:2012ut, Fong:2013hl}. At low temperatures, the absorbed photon energy in the graphene electrons can dissipate through three major channels: electronic heat diffusion, photon emission, and electron-phonon coupling. The electron heat diffusion is the heat transfer channel out of the graphene sheet to the electrodes. However, at the superconductor-graphene contact, Andreev reflection can suppress the thermal diffusion and quench this thermal conductance channel \cite{Fong:2013hl, Borzenets:2013vz}.

Photon emission from the graphene sheet to its EM environment can also be an effective cooling channel at low temperatures \cite{Schmidt:2004bg, Meschke:2006ce}. For a small measurement bandwidth $B$, such that $B<k_BT/h$ this radiation thermal conductance $G_{rad}$ is given by \ba G_{rad} \simeq r_0k_B B\ea where $r_0$ is the impedance matching factor. We can reduce this heat transfer channel by narrowing down the measurement bandwidth or deliberately mismatching the normal Jj resistance away from the amplifier input impedance. However, this may trade off the photon counting speed and Jj voltage measurement signal-to-noise ratio, respectively. For 1 MHz measurement bandwidth and $r_0 = 1$, $G_{rad} \simeq 1.4\times 10^{-17}$ W/K (Fig. 4b). This cooling channel is only significant at about 0.01 K when it is numerically comparable to the thermal conductance due to the electron-phonon coupling $G_{ep}$.

Internal energy can be transferred from electrons to phonons by scattering \cite{Viljas:2010jj, Chen:2012et}. Similar to Stefan-Boltzmann blackbody radiation, this heat transfer is a high power law in temperature, originating from the integral of bosonic and fermionic occupancies and density of states in the Fermi golden rule calculation. However, we can linearize this function to extract a thermal conductance $G_{ep}$.

The Bloch-Gr\"{u}neisen temperature $T_{BG} = 2\hbar s k_F/k_B$, where $s=26$ km s$^{-1}$ is the speed of sound in graphene, marks the temperature when the Fermi momentum of the electrons is comparable to that of the graphene acoustic phonons. For the carrier density we consider here, $T < T_{BG}$ and the heat transfer from electrons to acoustic phonons in graphene is given by \cite{Viljas:2010jj, Chen:2012et, Fong:2012ut, Betz:2012wya, Fong:2013hl, McKitterick:2016he}:
\begin{equation}
P_{ep}=\Sigma A(T_e^\delta-T_0^\delta)
\label{eqn:EPhHeatTransfer}\end{equation} where $\Sigma$ is the electron-phonon coupling parameter. The power $\delta$ is determined by the disorder in graphene. Disorder effects dominate the electron-phonon coupling when the typical phonon momentum is smaller than $1/l_{mfp}$, the inverse of the electron mean free path. Thus for $T$ higher (lower) than $T_{dis} = h s/k_B l_{mfp}$, the electron-phonon coupling is in the clean (disordered) limit. For clean graphene, $\delta$ = 4 and $\Sigma=\pi^{5/2}k_B^4\mathcal{D}^2n^{1/2}/(15\rho_m\hbar^4v_F^2s^3)$ whereas for disordered graphene, $\delta = 3$ and $\Sigma=2\zeta(3)k_B^3\mathcal{D}^2n^{1/2}/(\pi^{3/2}\rho_m\hbar^3v_F^2s^2l_{mfp})$, where $\mathcal{D}\simeq18$ eV is the deformation potential, $\rho_m=7.4\times10^{-19}$ kg $\mu$m$^{-2}$ is the mass density of graphene, and $\zeta$ is the Riemann zeta function. In the linear response regime, when $(T_e-T_0) \ll T_0$, the Fourier law is recovered: $P_{ep} \simeq G_{ep}(T_e - T_0)$ with $G_{ep} = \delta\Sigma A T_0^{\delta-1}$ as the electron-phonon thermal conductance (Fig. 4b). The total cooling power, including radiation and electron-phonon coupling, is $P_{ep}+G_{rad}(T_e-T_0)$. We can equate this rate of heat transfer to $C_e(dT/dt)$ such that
\begin{equation}
\Sigma A (T_e^\delta-T_0^\delta) + r_0 k_B B(T_e-T_0) \simeq -C_e\frac{dT_e}{dt}
\label{eqnTransient}\end{equation}In the linear response regime, Eqn. (\ref{eqnTransient}) reduces to a simple RC circuit with a thermal time constant $\tau_{th}=C_e/(G_{ep}+G_{rad})$ (Fig. 4c). Effectively, the weak electron-phonon coupling in graphene helps maintaining the heat in the electrons for a longer period of time. $\tau_{th}$ determines the intrinsic dead time of the graphene SPD so that an infrared detector operating at a few Kelvin could count photons at a rate up to GHz while a microwave detector operating at 10s of millikelvin could have a count rate in the MHz range. When $G_{ep}\gg G_{rad}$, thermal time constants increase rapidly as operating temperature decreases because $C_e \propto T$ while $G_{ep}$ has a higher temperature power law i.e. $T^{\delta-1}$ in monolayer graphene. In this regime $\tau_{th}$ is independent of carrier density because both $C_e$ and $G_{ep}$ are proportional to $n^{1/2}$. In contrast, when $G_{ep}\ll G_{rad}$, $\tau_{th}$ decreases with decreasing $T$ (Fig. 4c, green and orange curves) because $C_e$ decreases while $G_{rad}$ is constant in $T$ for a given bandwidth. $G_{rad}$ also has no $n$ dependence so that $\tau_{th}\propto \sqrt{n}$, allowing for the possibility to quickly reset the graphene SPD through its gate voltage.

We solve Eqn. (\ref{eqnTransient}) numerically to find $T(t)$ for the absorption of a single 26 GHz photon by a 1 $\mu$m$^2$ device operating at 25 mK and plot the result in Figure 4d. Because of the high-temperature power law in the electron-phonon heat transfer for $(T_e - T_0) \gg T_0$, $T_e$ drops faster than exponential, followed by a slow decay at the time constant $\tau_{th}$.

For this modeling, we employ $l_{mfp}$ = 120 nm at $n_0$, deduced from the electrical transport measurements on the typical gJj devices that we fabricate (See Section \ref{sectionGJJ}). This electrical transport mean-free-path would put the disorder temperature higher than the operating temperatures we consider here. Hereafter we will use the graphene thermal response in the disorder limit, i.e. $\delta = 3$.

\begin{table}[t]
\centering
\begin{tabular}{ l c c } 
\hline \hline
\multicolumn{3}{c}{\textbf{Modeling Parameters}}\\ 
\hline
Graphene dimensons& & 5 $\mu$m x 200 nm \\ 
Jj channel length&$L$& 200 nm\\ 
Jj channel width&$W$& 1.5 $\mu$m\\ 
Electron density&$n_0$& 1.67x10$^{12}$ cm$^{-2}$\\
Electronic mobility&$\mu$& 8000 cm$^2$/Vs\\
Jj normal resistance&$R_n$& 63 $\Omega$\\
Mean free path&$l_{mfp}$& 120 nm\\
Electronic heat capacity&$C_e$($T_0$)& 6.3 $k_B$\\
Disorder temperature&$T_{\rm{dis}}$& 10.4 K\\
Bloch-Gr\"{u}neisen temp.&$T_{BG}$& 90.5 K\\
$I_c(T_0)R_n$ product&$I_c(T_0)R_n$ & 223 $\mu$eV\\
Thouless energy&$E_{Th}$& 990 $\mu$eV\\
Jj coupling energy&$E_{J0}(T_0)$& 7.25 meV\\
Plasma Freq.&$\omega_{p0}(T_0)$& 156 GHz\\
McCumber parameters&$\beta_{SC}$&0.2\\
NbN superconducting gap&$\Delta_0$& 1.52 meV\\
\hline \hline
\end{tabular}
\label{tabPara}
\caption{List of device parameters to model the graphene-based Josephson junction single photon detector in this report.}
\end{table}

\section{Graphene-based Josephson junction}
\label{sectionGJJ}
Detection of single photons relies on the gJj transition from the zero-voltage to resistive state. This switching is a probabilistic process described by the escape rate, $\Gamma$, of the Jj phase particle from the tilted washboard potential in the resistively and capacitively shunted junction (RCSJ) model \cite{Tinkham}. The rate of switching depends intimately on the Jj critical current, $I_c$, which takes on different forms as a function of temperature depending on whether the Jj is short or long and whether it is diffusive or ballistic. Superconductor-graphene-superconductor Jjs have been studied in these different regimes \cite{Lee:2011et, Jeong:2011dp, Mizuno:2013dz, Calado:2015fp, BenShalom:2015hy, Borzenets:2016gw}. However, experimental values, such as $I_c$ and the parameters in the long diffusive junction, have fallen short of theoretical expectations due to impurity doping. To emphasize the feasibility of gJj SPDs under currently realizable parameters, we model the device performance based on experimentally determined values (summarized in Table 1) of the gJj shown in the inset of Fig. 5b instead of analyzing the gJj from a purely theoretical standpoint.

\begin{figure}[t]
\includegraphics[width=\columnwidth]{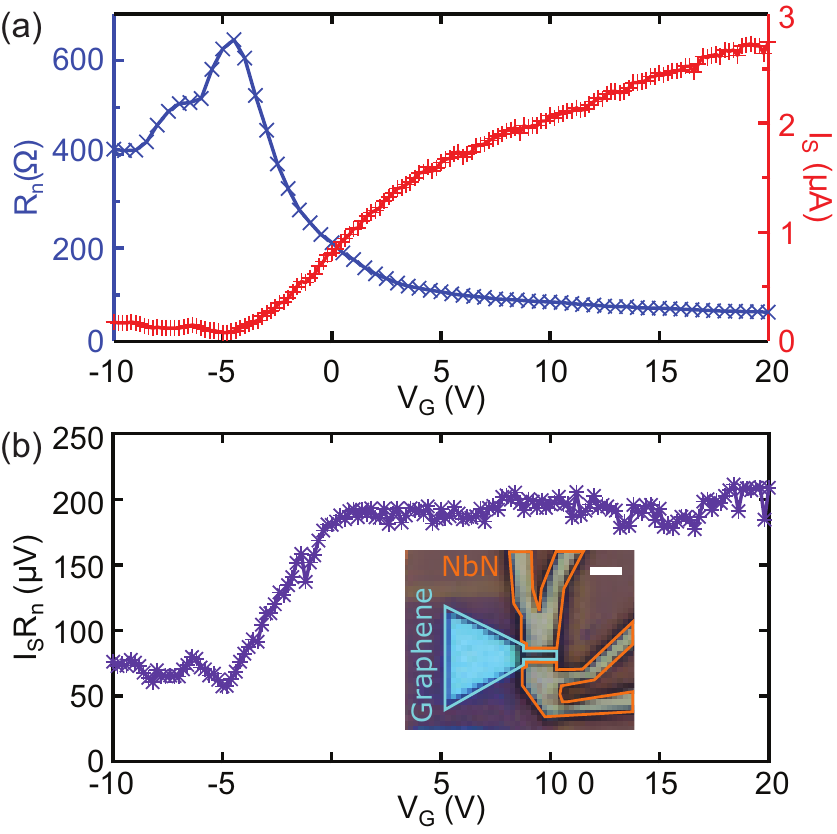}
\caption{(a) Measured $R_n$ (blue) and $I_s$ (red) as a function of $V_G$ (b) $I_sR_n$ versus $V_G$. Inset: Optical micrograph of the measured gJj whose device performance parameters were used in this report. Blue colored region is the graphene channel encapsulated by hBN and the region emphasized by orange colored lines are the NbN electrodes. Scale bar (white) is 1.5 $\mu$m.}
\label{figgJJprops}
\end{figure}

The measured graphene monolayer is encapsulated between atomically flat and insulating boron nitride ($\sim$30 nm thick) using a dry-transfer technique \cite{Wang:2013ch}. The doped silicon substrate serves as a back gate electrode. The superconducting terminals consist of 5 nm thick niobium and 60 nm thick niobium nitride (NbN) after reactive ion etch and electron beam deposition of 5 nm titanium to form the etched one-dimensional contact \cite{Lee:2016uv}. The distance between the superconducting electrodes $L$ is about 200 nm, forming a proximitized Jj with monolayer graphene as the weak link.

The gJj is mounted at the mixing chamber (MC) of a dilution refrigerator with base temperature 25 mK. The electrical transport measurement is performed through a standard four-terminal configuration with the silicon substrate as the back gate to control the carrier density in graphene. All DC measurement wires are filtered by two-stage low pass RC filter mounted at the MC with an 8 kHz cutoff frequency. $I_b$ is set by a DC voltage output through a 1 M$\Omega$ resistor while the voltage measurements are taken by a data acquisition board after a low noise preamplifier with a 10 kHz low pass filter.

Fig. 5a shows the resistance as a function of gate voltage. CNP is observed to occur at $V_G=-5$V. The chosen operating carrier density $n_0$ corresponds to $V_G=20$ V for a 300 nm thick dielectric material composed of silicon dioxide and hexagonal boron nitride. At this gate voltage $R_n$ is measured to be 63 $\Omega$. The mobility is calculated as $\mu = L/(neR_nW)$, the mean free path as $l_{mfp} = \hbar\mu(\pi n)^{1/2}/e$, and the diffusion coefficient as $D_e = v_Fl_{mfp}/2$. At $n_0$ this yields $\mu =8000$ cm$^2$V$^{-1}$s$^{-1}$, $l_{mfp} = 120$ nm, and $D_e = .06$ m$^2$s$^{-1}$.The slope of conductance, $\sigma=L/(R_nW)$, versus $n$ gives similar results for mobility using $\mu=\frac{1}{e}\frac{d\sigma}{dn}$. $l_{mfp}\simeq120$ nm is comparable to the channel length $L\simeq$ 200 nm so this device is in neither purely ballistic nor purely diffusive regime.
When the bias current through the Jj increases, it switches from supercurrent to resistive state state at a switching current $I_s$ depending on the gate voltage as shown in Fig. 5a. Despite $I_s$ and $R_n$ vary for different gate voltages, their product approaches to a constant in both the highly electron- and hole-doped regime (Fig. 5b). Consistent to the experimental results in Ref. \cite{Jeong:2011dp, Mizuno:2013dz, BenShalom:2015hy}, the $I_sR_n$ product for the niobium-based gJj has a smaller value than the calculation from the superconductor critical temperature probably due to impurity doping.

Fig. 6a shows typical IV characteristics for $V_G=20$V at 25 mK and 200 mK, which correspond to $T_0$ and $T_{peak}$ at 26 GHz, respectively (Fig. 4d), measured by ramping up the bias current at a rate of 0.1 $\mu$A/s. 
In order to measure the escape rate, we repeat the IV measurement 100 times at each base temperature and find the average switching current, $\langle I_s(T)\rangle$ (Fig. 6a inset), which at this sweep rate is typically about 90\% of $I_c$. From the histogram of these 100 gJj switching events, we obtain the probability density of the switching current $P(I_s)$ for each measured temperature. Fig. 6c shows $P(I_s)$ at 25 mK and 200 mK. We can derive the phase particle escape rate $\Gamma$ from the $P(I_s)$ data using \cite{Fulton:1974uf}: \ba P(I_s) = \left[\Gamma(I_s)/\left(\frac{dI}{dt}\right)\right]\left(1-\int_0^{I_s}P(I^\prime)dI^\prime\right)\ea where $dI/dt$ is the bias current ramping speed.

\begin{figure}[t]
\includegraphics[width=\columnwidth]{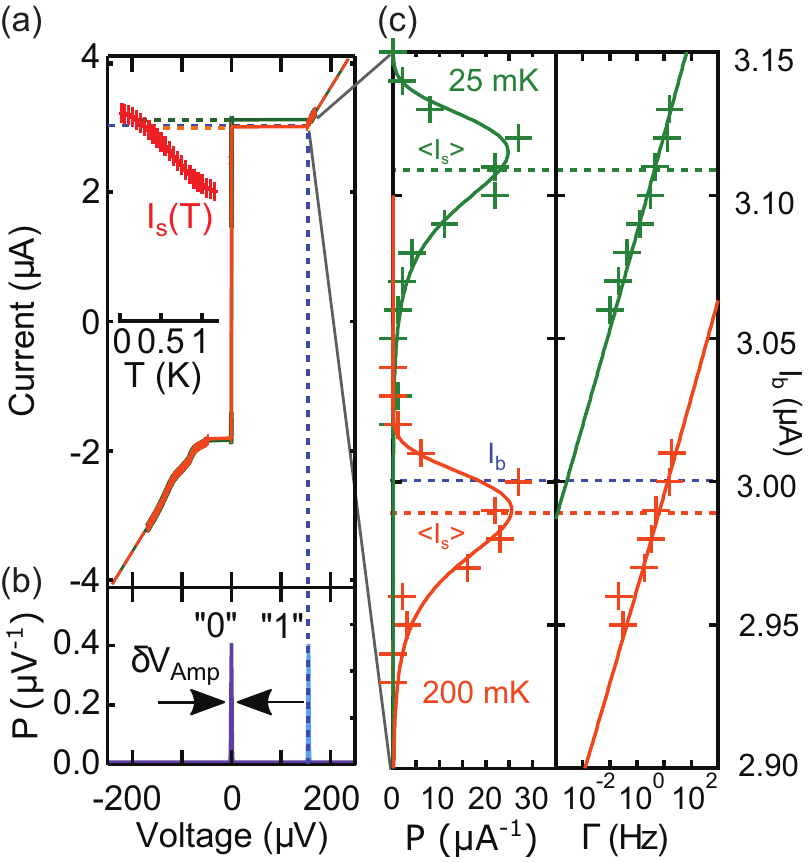}
\caption{(a) Measured gJj $IV$ characteristics for electron temperature at $T_0$ = 25 mK (green) and $T_{peak}$ = 200 mK (orange) with carrier density of $1.7\times10^{12}$ cm$^{-2}$. Inset: The measured average switching current vs. temperature. (b) The Jj voltage and the expected voltage noise from an amplifier at 1 MHz bandwidth. The blue dotted line depicts, for a given bias current in a photon event, the order of magnitude of the Jj voltage $V=I_bR_n$. (c) The switching probability and escape rate of the phase particle as a function of the Jj current bias. The solid line is the best fit probability distribution to the data assuming the escape mechanism is MQT. Solid lines are given by Eqn. 9 with $I_c$ = 3.565 $\mu$A (25 mK) and $I_c$ = 3.425 $\mu$A (200 mK). }
\label{figJJ}
\end{figure}

Fig. 5C right panel shows the extracted $\Gamma$ data which can be described by \cite{Devoret:1985jx}: \ba \Gamma = A\exp{\left(-\frac{\Delta U}{k_BT_{esc}}\right)}\ea where $\Delta U = 2E_{J0}(\sqrt{1-\gamma_{JJ}^2}-\gamma_{JJ}\cos^{-1}{\gamma_{JJ}})$ is the energy barrier of the washboard potential and $T_{esc}$ is the ``escape temperature" which sets the energy scale competing with $\Delta U$ for the phase particle to escape from the washboard potential. Here, $E_{J0} = \hbar I_c/(2e)$ is the Josephson coupling energy and $\gamma_{Jj}=I_b/I_c$ is the normalized bias current. In the thermal activation (TA) regime, $T_{esc}$ is simply the temperature of the device while in the macroscopic quantum tunneling (MQT) regime
\ba T_{esc}=\hbar \omega_p/\left[7.2k_B\left(1+\frac{0.87}{Q}\right)\right]\ea where $Q=\omega_pR_nC_{Jj}$ is the Jj quality factor with $\omega_p = \omega_{p0}(1-\gamma_{Jj}^2)^{1/4}$ being the Jj plasma frequency, $\omega_{p0} = (2eI_c/(\hbar C_{Jj}))^{1/2}$ being the zero bias Jj plasma frequency, and $C_{Jj}$ being the effective junction capacitance. While the geometrical capacitance between the superconducting terminals is estimated to be sub-femtofarad lower bounded by the parasitic capacitance to the substrate, there is an effective capacitance due to electronic diffusion given by $C_{Jj}=\hbar/R_nE_{Th}$ \cite{Lee:2011et}. $E_{Th}= \hbar D_e/L^2$ is the Thouless energy \cite{Dubos:2001ca} where $D_e=v_Fl_{mfp}/2$ is the diffusion constant. $E_{Th}$ is the characteristic energy scale of a diffusive gJj and is estimated to be $\sim$11 K for the device characterized for our modeling which gives $C_{Jj}\sim$11 fF. Using the effective capacitance, we estimate $\omega_{p0}$ of this gJj to be $2\pi\times156$ GHz. The Stewart-McCumber parameter $\beta_{SC} = Q^2$ is about 0.2. This implies that the gJj should be an overdamped junction with its phase particle retrapped quickly after switching to the resistive state. However, the measured junction is hysteretic probably because the resistive state bias current self-heats the junction at low temperatures \cite{Borzenets:2013vz}.

Using $\omega_p$ and $Q$ above, we can estimate this gJj to be in the MQT regime for temperatures below 470 mK \cite{Lee:2011et}. With the prefactor $A$ of Eqn. 9 given by 
\ba
A=\begin{cases}
\frac{\omega_p}{2\pi}\left(\sqrt{1+\frac{1}{4Q^2}}-\frac{1}{2Q}\right) &\text{(TA)}\\
12\omega_p\sqrt{\frac{3\Delta U}{2\pi \hbar \omega_p}} &\text{(MQT)}
\end{cases}
\ea 
we fit the extracted $\Gamma$ values with $I_c$ as the only fitting parameter (all other parameters depending on $R_n$, $C_{Jj}$, and $I_b$) and find the fitting consistent with the MQT process as expected. The solid lines in Fig. 6c show the MQT $\Gamma$ with $I_c$ equal to 3.57 $\mu$A and 3.43 $\mu$A at 25 mK and 200 mK, respectively. The experimentally determined $\Gamma$ as a function of $I_b$ and $I_c(T)$ will help us to calculate the SPD performance in Section V.

\section{Photon detection performance}\label{sectionPerformance}

The gJj can be set to detect single photons by setting a bias current $I_b$ (as an example, $I_b \simeq 3.03\mu$A in Fig. 6c) below the switching current, so that when photons raise the graphene electron temperature, the gJj may switch to the resistive state as its critical current quenches. As the electrons cool, the gJj will switch back to the superconducting state for a junction with no hysteresis. For a hysteretic junction, the detector can be reinitialized using the bias current or gate voltage, as both the switching and retrapping current depend on the gate voltage. The gJj can function as an SPD when either the MQT or the TA process dominate the Jj transition because its operation depends on the change of $\Gamma$ as the electron temperature increases. However, the phase diffusion process \cite{Lee:2011et} is not desirable because the finite sub-gap resistance can diminish the signal-to-noise ratio of the gJj voltage signal readout significantly.

The voltage drop across the gJj upon detection is given by $I_bR_n$ which is about 190 $\mu$V for the measured device (marked by the vertical dashed line in Fig. 6a). The inaccuracy in measuring this voltage with a short averaging time will be dominated by the amplifier noise because the Johnson noise spectral density, $\sqrt{4R_nk_BT}$, is only about 10 pV/Hz$^{1/2}$ while the shot noise, $ R_n \sqrt{\mathcal{F}2eI_b}$, is about 33 pV/Hz$^{1/2}$ for $I_b = 3.03 \mu A$ and Fano factor $\mathcal{F} \simeq 0.29$ \cite{DiCarlo:2008iv, Danneau:2008p196802}. Using a typical low noise voltage amplifier with power spectral density of 1 nV/Hz$^{1/2}$, the amplifier voltage noise at 1 MHz measurement bandwidth is 1 $\mu$V. Signal-to-noise of the gJj readout is large as depicted in Fig. 6c, where two well separated Gaussian peaks centered at $V_b=0$ (``0'' no photon state) and $V_b=I_bR_n$ = 190 $\mu$V (``1'' photon state) with FWHM corresponding to the amplifier voltage noise. A Josephson coupling energy that gives an $I_cR_n$ product of about 100 $\mu$V will be sufficient for making a gJj SPD.

\begin{figure*}[t]
\includegraphics[width=2\columnwidth]{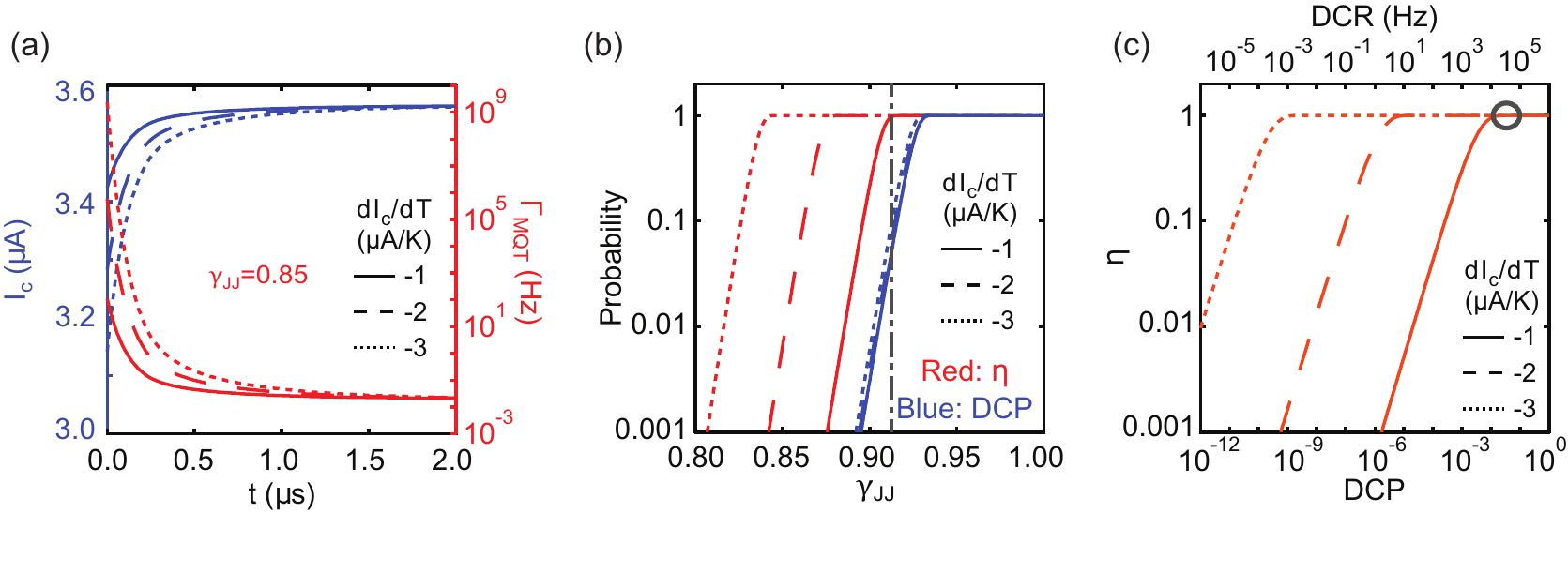}
\caption{(a) Change of critical current and phase slipping probability due to a single microwave photon, impinging at $t$ = 0 s and $T$ = 0.025 K, deduced from the graphene thermal properties and measured gJj parameters. (b) Intrinsic efficiency and dark count probability at 1 MHz count rate versus bias current. Vertical gray dashed line indicates bias current for which $\eta>$0.99 for the measured $dI_c/dT$. (c) Trade-off between the intrinsic quantum efficiency $\eta$ and the dark count rate at various bias currents from (b). Gray circle corresponds to gray line from (b).}
\label{figPerformance}
\end{figure*}

The performance of a gJj SPD can be calculated by the probabilistic Jj transition at an elevated temperature. We choose to benchmark the gJj SPD by its intrinsic quantum efficiency $\eta$ and dark count probability $\mathcal{P}_{dark}$ for single photon detection at 26 GHz with improved performance expected for higher energy photons. We note that for infrared photons, the peak temperature will be higher than the superconducting gap energy of niobium nitride, resulting in heat leakage that can reduce the peak temperature and shorten the duration of the heat pulse. Calculation of the spectral current \cite{Wilhelm:1998hk} will probably be required to understand the gJj under high energy photon excitation and is beyond the scope of this work. For microwave photons, the temperature rise is much smaller than the gap energy and the detector remains in the quasi-equilibrium regime.

We can calculate $\mathcal{P}_{dark}$ using $\Gamma$. The total escape probability in a measurement integration time $t_{meas}$ is given by $1-\exp(-\int_0^{t_{meas}} \Gamma(\tau) d\tau)$. In the absence of incident photons, $\Gamma $ is equal to the dark count rate $\Gamma_{dark} $  and would be a constant $\Gamma_0(T=T_0)$ if the temperature were fluctuation free. To include the effect of temperature fluctuation, we use the averaged $\Gamma$, i.e. $\langle\Gamma\rangle=\int_0^\infty dT p(T) \Gamma(I_c(T))$ where $p(T)$ is a Gaussian distribution centered at $T_0$ with standard deviation $\Delta T$. Since $\Gamma$ increases quickly as a function of $T$, $\langle\Gamma\rangle\geq\Gamma_0$. $\mathcal{P}_{dark}$ equals $1-\exp(-\langle\Gamma\rangle t_{meas})$ and can be significantly higher than $1-\exp(-\Gamma_0t_{meas})$ depending on the size of $dI_c/dT$ and $\Delta T/T_0$. At $T_0$ = 25 mK and $I_b=$3.28 $\mu$A ($\gamma_{Jj}\simeq0.91$), $\mathcal{P}_{dark}$ for $t_{meas}$ = 1 $\mu$s is about 0.07 using an estimated $\Delta T/T_0$ = 0.8 (Fig. 3C) and the fitted $\Gamma$ from Section IV.

Upon photon absorption, the electron gas heating and the subsequent cooling from $T_{peak}$ result in a time-dependent $\Gamma$ that can be used to calculate $\eta$. Using the $T_e(t)$ in Fig. 4d, $I_s(T)$ in Fig. 6a, and fitted $\Gamma(I_c)$, we calculate both the critical current and $\Gamma$ as they recover to their nominal values after the photon incidences at $t=0$ (see Fig. 7a). We calculate the intrinsic quantum efficiency $\eta$ of the gJj SPD from \ba\eta = 1-\exp\left(-\int_0^{t_{meas}} \Gamma\left(I_c(\tau)\right) d\tau\right)\label{EqnEta}.\ea The detection efficiency increases with the measurement time, but so does $\mathcal{P}_{dark}$. To balance between these competing effects, we benchmark the gJj SPD performance by taking the measurement integration time to be 1 $\mu$s such that $\Gamma(I_c(t_{meas}))\simeq2\langle\Gamma\rangle_{T=T_0}$. Fig. 7b plots $\eta$ and $\mathcal{P}_{dark}$ as a function of $\gamma_{Jj}$ for three different $dI_c/dT$ values. With the $dI_c/dT$ of the measured gJj at -1.1 $\mu$A/K (measured in the linear region of $\langle I_s(T)\rangle$ above about 100 mK in Fig. 6a), we can set $\gamma_{Jj}$ = 0.91 (vertical grey dashed line) to reach an intrinsic quantum efficiency $>$0.99 while maintaining $\mathcal{P}_{dark}$ $\simeq$ 0.07.

We plot the trade off between $\eta$ and $\mathcal{P}_{dark}$ in Fig. 7c by eliminating the common parameter $\gamma_{Jj}$ of Fig. 7b. Favorable SPD performance occurs in the plateau region where $\eta$ approaches unity while $\mathcal{P}_{dark}$ remains small. This SPD regime is feasible within the parameters of existing gJjs that we can fabricate. The operating point circled in Fig. 7c corresponds to the same bias current and integration time setting at the vertical dashed lines in Fig. 7b.

To optimize the SPD performance or to operate at a different frequency, we argue that the design should focus on the dependence of the gJj critical current on temperature. Although a lower operating temperature and a smaller area of monolayer graphene can enhance the temperature rise due to a smaller electronic heat capacity, the rate of improvement can quickly diminish because the heat loss from the electrons is dominated by the coupling to phonons and is faster than exponential when $T_e - T_0 > T_0$. The time integral of $\Gamma$ in Eqn. (11) can only increase marginally to enhance the intrinsic quantum efficiency insignificantly. However, the phase particle escape rate of the Jj can increase by orders of magnitude with increased $dI_c/dT$ because of the exponential dependence of $\Delta U$ in both MQT and TA processes, as suggested by the initial numerical values of $\Gamma$ in Fig. 7a. This behavior contributes drastically to the SPD intrinsic efficiency, allowing a lower Jj current bias to further reduce the dark count probability as shown in Figs. 6b and 6c by the dashed and dotted lines for double and triple the measured $dI_c/dT$, respectively. Merely doubling $dI_c/dT$ lowers $\mathcal{P}_{dark}$corresponding to $\eta=$0.99 by 5 orders of magnitude while an 8 order of magnitude improvement is possible by tripling $dI_c/dT$. The operation of a gJj SPD at a higher temperature up to 4 K is possible using a ballistic-like gJj such as that demonstrated in Ref. \cite{Borzenets:2016gw} with $dI_c/dT$ on the order of -1 $\mu$A/K.

\section{conclusion}
In conclusion, we have introduced a device scheme for ultra-broadband single photon detection based on the extremely small electronic specific heat in graphene, which results from the vanishingly small electronic density of states near the Dirac point and its linear band structure. Our model analysis shows that single photon detection, together with very small dark count rates, is possible across a wide spectral region, from the microwave to near-infrared light. Efficient light absorption into the graphene absorber can be achieved by impedance matching structures, such as metallic or dielectric resonators considered here for microwave and optical frequencies. Using experimental parameters from a fabricated device, we model that a gJj SPD operating at 25 mK could reach a system detection efficiency higher than 99\%, together with one-tenth dark count probability for a 26 GHz photon. This device performance should improve for higher photon energies. Inductive readout \cite{Giazotto:2008, Voutilainen:2010, Gasparinetti:2015gr, Govenius:2016} can be used in the future to increase the SPD operation bandwidth by avoiding the Joule heating when the gJj switches to resistive state. We have only explored a small set of possible parameters in this report. Further optimization of the gJj SPD will depend on application-specific performance trade-offs. Heat leakage to the superconducting electrodes as well as position and time dependence of the heat propagation will need to be included in the future for a more realistic prediction and a better understanding of the fundamental limits of the gJj SPD. The rapid progress in integrating graphene and other Van der Waals materials with established electronics platforms, such as CMOS chips, provides a promising path towards single photon-resolving imaging arrays, quantum information processing applications of optical and microwave photons, and other applications benefiting from quantum-limited photon detection of low-energy photons.

\textbf{Acknowledgements.} We thank valuable discussions with L. Levitov, S. Guha, B.-I. Wu, J. Habif, and M. Soltani. E.D.W. was supported in part by the Office of Naval Research (N00014-14-1-0349). The work at Harvard was supported by the Science and Technology Center for Integrated Quantum Materials, NSF Grant No. DMR-1231319. Numerical simulations were supported in part by the Center for Excitonics, Center for Excitonics, an Energy Frontier Research Center funded by the U.S. Department of Energy, Office of Science, Office of Basic Energy Sciences under award no. DE-SC0001088. The works of T.A.O. and K.C.F. were funded by the Internal Research and Development in Raytheon BBN Technologies.

\bibliographystyle{apsrev}

\end{document}